\documentclass[a4paper]{article}
\usepackage{hyperref}
\usepackage{amsmath}
\usepackage{siunitx}
\usepackage{amsfonts}
\usepackage{amssymb}
\usepackage{amsmath}
\usepackage{graphicx}
\usepackage{caption}
\usepackage{subcaption}
\usepackage[english]{babel}
\usepackage{gensymb}
\usepackage{array}
\usepackage{xfrac}
\usepackage{upgreek}
\usepackage{color, soul}
\usepackage[toc,page]{appendix}
\usepackage{fullpage}
\usepackage[square, numbers, super]{natbib}
\numberwithin{equation}{section}
\begin{document}
\title{Information Mechanics}
\author{Kiyam Lin \& SongLing Lin}
\date{\today}
\maketitle

\begin{abstract}
Despite the wide usage of information as a concept in science, we have yet to develop a clear \& concise scientific definition. This paper is aimed at laying the foundations for a new theory concerning the mechanics of information alongside its intimate relationship with working processes. Principally it aims to provide a better understanding of what information is. We find that like entropy, information is also a state variable, and both their values are surprisingly contextual. Conversely, contrary to popular belief, we find that information is not negative entropy. However, unlike entropy, information can be both positive and negative. We further find that it is possible to consider a communications process as a working process and that Shannon's entropy is only applicable to the modelling of statistical distributions. In extension, it appears that information could exist within any system, even thermodynamic systems at equilibrium. Surprisingly, the amount of mechanical work a thermodynamic system can do directly relates to the appropriate corresponding information available. If the system does not have corresponding information, it can do no work irrespective of how much energy it may contain. Following the theory we present, it will become evident to the reader that information is an intrinsic property which exists naturally within our universe, and not an abstract human notion.

\end{abstract}

Since the publication of Shannon's pioneering work on information theory \cite{shannon48}, information has increasingly been viewed as a scientific concept \cite{lombardi04} \cite{brier05} \cite{logan14}, because we have obtained a method of measuring it. However Shannon's method of measuring information was not easily adopted by many different disciplines, as it does not necessarily relate to the context fittingly \cite{lombardi04} \cite{brier05} \cite{sloman11}. Although information is generally agreed upon to affect various processes, different disciplines view information from different perspectives \cite{lombardi04} \cite{sloman11} \cite{smith00}, as their usage of information differs depending on their respective working fields. This has resulted in there being many different definitions of information \cite{smith00} \cite{sloman11}, and due to these complications, information has still not become a fundamental scientific concept.

If we look back to Shannon's work, a key concept he introduced was to relate the gain of information to the reduction of entropy. Similarly, the concept of entropy came about from the analysis of doing work in a thermodynamic system, and Boltzmann showed us that entropy arises due to the uncertainty in initial conditions in the complex micro-states of thermodynamic systems. These ideas suggest to us that maybe information could also be related to doing work in thermodynamic systems. Taking this a step further, we may hypothesize that information not only affects thermodynamic working processes, but also any working process that could contain uncertain factors. The following research explores the nature of information by analysing its affect on working processes.
\section{Information to Conduct a Specific Work}
From our understanding of physics we know that to be able to do work, we need energy. In real world situations however, we are usually unable to achieve 100\% energy efficiency. If we closely analyse working processes in various systems, we realise that unwanted outcomes are inevitably produced, which lower the working efficiency. This is due to almost every work having the potential to produce multiple outcomes, of which only some are wanted outcomes, fulfilling our working goal(s). When we cannot control all of the factors affecting our working process, only part of our work would effectively be successfully utilised in producing the wanted outcomes, and the rest would produce unwanted outcomes. Therefore we may realise, how well we control our working process decides the working efficiency.

For example, a photographer wishes to take photographs of a landing Airbus A380 at Heathrow airport. Due to there being many different planes landing at Heathrow, if the photographer is unable to determine before taking each shot if a plane is an Airbus A380 or not, he would have to randomly decide whether he should take a photo. Naturally the probability of him taking the photo he wants per shot is low, resulting in a low overall working efficiency. However, suppose the photographer knows that Airbus A380's have 4 engines, and is able to accurately determine the number of engines a plane has during its descent. If he uses this information to help him decide whether he should take a shot or not, the probability of him taking the correct photo per shot would naturally increase, along with his overall working efficiency. It seems the more information gained, the better the control of the working process, and the higher the working efficiency.
 
\begin{figure}[h]
	\captionsetup{width=0.75\textwidth}
	\centering
	\includegraphics[width=0.50\textwidth]{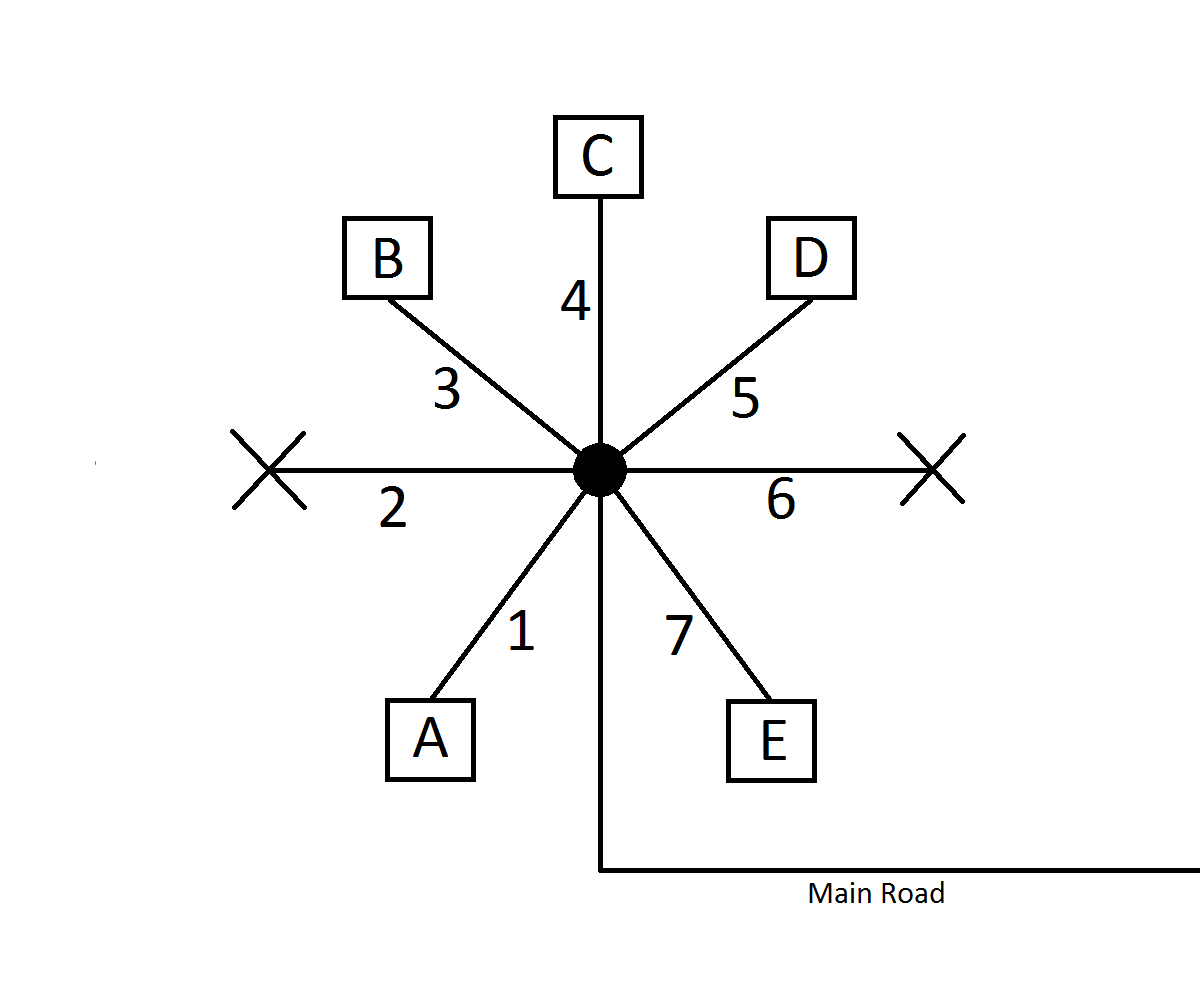}
	\caption{A diagram of Village X, after the courier enters the village from the main road, there are 7 possible routes. There are 5 residencies, A, B, C, D \& E with roads leading to them as labelled, and roads 2 \& 6 lead to no residencies.}
	\label{fig:f1}
\end{figure}

To further explore the relationship between information and the process of doing a specific work, we may analyse an idealised example of a courier, named Alan, delivering a parcel to a specific residency in village X. A diagram of village X is depicted in figure \ref{fig:f1}.

Suppose Alan has a parcel to deliver to B. After entering village X, Alan has a choice of 7 roads. If we consider embarking on a road once as a single working trial, and a trial on each different road a separate working category, there are 7 working categories in this scenario. Without information to help him identify which road leads to B, he would have to conduct work by random trial (randomly choosing a road and seeing if it leads to B). His probability to achieve the working goal in one trial would only be $\frac{1}{7}$. We can see that he would have a fairly low working efficiency, producing some unwanted outcomes before he achieves the working goal.
 
Suppose Alan now has information that lets him know that roads 2 and 6 lead to nowhere. He would avoid choosing these roads in his future work. If he still chooses the other roads at random, his probability of achieving the working goal in one trial would increase to $\frac{1}{5}$, a higher working efficiency.

Suppose Alan now has information that lets him know specifically that road 3 leads to B, this is equivalent to knowing that roads 1, 2, 4, 5, 6 and 7 do not lead to B, so he avoids choosing those roads. His probability of achieving the working goal in one trial would increase further to 1, achieving 100\% working efficiency.

From this we can see that the working efficiency is affected by the probability of achieving the working goal in one trial. It seems that in a system with a fixed number of working categories, how much information Alan has directly decides the probability of achieving the working goal in one trial. More generally, we may quantify information using the following equation.

\begin{equation} \label{eq:1infodef}
I = k \log(\Theta \cdot p)
\end{equation}

\noindent where $p$ is the probability of success (probability of achieving the working goal in one trial), $\Theta$ is the number of working categories, and $k$ is an arbitrary constant of proportionality. $I$ denotes the information to direct a specific work. The value that $k$ takes on is dependent on what units we wish to work in. For example, if we use $\log$ of base 2 and wish for $bits$ we would take $k = 1$. Yet, working in the same logarithmic base, and we wished for units of $nats$, we would take $k = \frac{1}{\log_{2} e}$.

From the above, we see that in a working system with a fixed number of working categories, the more information we have, the greater the probability of success. If we wish to attain the same probability of success in a working system with a greater number of working categories, more information is needed.

Although it appears that we need information when we do any kind of complicated work, we may notice that it is not consumed upon its use, unlike energy. This lets us understand that unlike energy, which is a necessity for doing work, without information, work can still be done. The possible working categories, and the amount of energy decides which processes can occur, however information only affects working trends. We may comprehend from this that information does not have an equivalent meaning to energy. But if we recognise energy do be our capacity to do work, we may see that information is more likely to be the capacity to affect working trends.

\section{Information to Conduct a Generic Work in a System}
Thinking about the example proposed in Chapter 1, Alan, as a courier, may not only have to do the specific work of delivering a parcel to B once, but also deliver different parcels to all residencies A, B, C, D or E repeatedly. From figure \ref{fig:f1}, we can see that the specific works of delivering a parcel to A, B, C, D or E are of the same type, and share the same working categories. Considering the generic work Alan does in village X, there are 7 working categories, but only 5 various working goals relating to different specific works.

If Alan has no information, and can only conduct his works by random trial, then his probability of success would be $\frac{1}{7}$ for all residencies. Similarly if Alan knew which road lead to which destination, i.e. has complete information for each specific work, then his probability of success for each specific work would be 1.

However, more often than not, Alan may have differing amounts of information to conduct each specific work, this may be caused by differences in route familiarity, working experience etc. What this would imply is a different average probability of success for different specific works. If we wish to consider Alan doing a generic work in village X, the amount of information Alan has to direct a generic work should be an average of the information he has to direct each specific work. If statistically, the probability of a parcel being intended for each residency differs, then the average taken must be a weighted average.  Suppose the amount of information Alan has to direct each specific work is $(I_{A} , I_{B} , I_{C} , I_{D} , I_{E})$, and the probability of occurrence of each working goal is $(q_{A} , q_{B} , q_{C} , q_{D} , q_{E})$, the amount of information to direct a generic work would be $I = q_{A} I_{A} + q_{B} I_{B} + \ldots + q_{E} I_{E}$.

We may deduce that for a general working system, if the probability of success for each working goal is $(p_{1} , p_{2} , \ldots , p_{N})$, and the probability of occurrence of each working goal is $(q_{1} , q_{2} , \ldots , q_{N})$ and $\sum_{i=1}^{N} q_{i} = 1$, according to equation (\ref{eq:1infodef}), the amount of information to direct a generic work would be given as

\begin{equation} \label{eq:2geninfo}
I = k \sum_{i=1}^{N} q_{i} \log (\Theta \cdot p_{i})
\end{equation}

\noindent where $\Theta$ is the number of working categories, $N$ is the number of working goals, $q_{i}$ is the probability of occurrence of a working goal, and $p_{i}$ is the probability of success for a specific working goal. $I$ denotes the information to direct a generic work, and $k$ is a constant of proportionality.

Equation (\ref{eq:2geninfo}) can be re-arranged into

\begin{equation} \label{eq:2geninfoexp}
I = k \log(\Theta) - (- k \sum_{i=1}^{N} q_{i} \log p_{i})
\end{equation}

\noindent we can see that the first term on the right hand side of the equation is only related to the number of working categories, $\Theta$, a variable only defined by the circumstances of the working system, independent of the state of the worker. Therefore, we may define this as the working diversity of the system, giving us

\begin{equation} \label{eq:2divdef}
D = k \log \Theta
\end{equation}

\noindent where $D$ is the working diversity, $k$ is a constant of proportionality and $\Theta$ is the number of working categories. From this equation, we may understand that working diversity represents the size of the working domain. The greater the size of the working domain, the more possible working categories.

The second term on the right hand side of equation (\ref{eq:2geninfoexp}) is only related to the probabilities of success and probabilities of occurrence. From the above, we know that the less information there is to direct any specific work, the lower the corresponding probability of success, similarly, for a generic work, less information would relate to lower probabilities of success of conducting generic works. This would be reflected in a greater uncertainty of achieving the working goal in each trial. Therefore we may define this term as the working entropy concerning a generic work

\begin{equation} \label{eq:2endef}
S = - k \sum_{i=1}^{N} q_{i} \log p_{i}
\end{equation}

\noindent where $S$ is the working entropy. $N$ is the number of working goals, $q_{i}$ is the probability of occurrence of any working goal, $p_{i}$ is the probability of success for a specific working goal, and $k$ is a constant of proportionality. However, in this equation, $p_{i}$ is generally independent of $q_{i}$, and whereas $q_{i}$ is constrained by $\sum q_{i} = 1$, there is no such constraint for $p_{i}$. Similarly, the value of $p_{i}$ associated with each particular working goal is not collectively constrained, and each $p_{i}$ can independently have any value in the range $0 \leq p_{i} \leq 1$. When we conduct generic works, the more uncertain we are in achieving the working goal, the more working entropy there is.

Thus we arrive at

\begin{equation} \label{eq:2IDS}
I = D - S
\end{equation}

This equation shows us a simple relationship between information, diversity and entropy on the meaning of conducting work. However, as $D$ and $S$ are independent variables, we also have

\begin{equation} \label{eq:2deltaIDS}
\Delta I = \Delta D - \Delta S
\end{equation}

This strongly suggests to us that information is not negative entropy, and only when the quantity of working diversity is constant, does an increase in information lead to a direct decrease in entropy, or vice versa. However, a system with a changing amount of diversity would result in a more complex relationship between information and entropy. 

Looking back to the previous section, we find that equation (\ref{eq:1infodef}) can be re-arranged into the same form as equation (\ref{eq:2IDS}), meaning the concepts of working entropy and diversity also apply to any specific work.
\section{Positive and Negative Information}
Looking at equation (\ref{eq:2endef}), as $p_{i}$ can take any value between $0 \leq p_{i} \leq 1$, $S$ would have a corresponding range of $0 \leq S \leq \infty$. However, for any working system with a finite number of working categories, according to equation (\ref{eq:2divdef}), the value of the working diversity must also be finite. With reference to equation (\ref{eq:2IDS}), we may see that $I$ can both be $I \geq 0$ or $I < 0$. This seemingly means that information can be negative.

To explore the meaning of negative information, let us think back to the example given in the first section, when Alan had no information to direct his work of delivering a parcel to B. Alan was only able conduct his work by random trial and his probability of success was $\frac{1}{7}$, according to equation (\ref{eq:1infodef}) Alan's work was directed by $0$ amount of information. Suppose Alan has some information that influences how he does his work. If this information leads him to believe that road 3 has a 90\% chance of leading to B, the probability of road 3 being trialled on each working attempt would be 90\%. In this scenario, we may see that his probability of success is $\frac{9}{10}$, and Alan has $2.66$ $bits$ of information. However, suppose instead the information he has led him to believe that road 3 had a 10\% chance of leading to B, the probability of road 3 being trialled on each working attempt would be 10\%. In this second scenario, we may see that his probability of success would be $\frac{1}{10}$ and Alan has $-0.51$ $bits$ of information.

From the above analysis, we saw that when Alan had a positive amount of information, his probability of success was greater than when he had no information. Conversely, when Alan had a negative amount of information, his probability of success was less than when he had no information. In a more general sense, for a specific work, according to equation (\ref{eq:1infodef}), when $p = \frac{1}{\Theta}$, $I = 0$. Similarly, if  $p > \frac{1}{\Theta}$, $I > 0$, whereas if $p < \frac{1}{\Theta}$, $I < 0$. 

For a generic work, if we suppose there is an average probability of success, according to equation (\ref{eq:2endef}) we have

\begin{equation} \label{eq:3avgenen}
S = - k \log \bar{p} = - k \sum_{i=1}^{N} q_{i} \log p_{i}
\end{equation}

\noindent where $\bar{p}$ is the average probability of success, which gives us

\begin{equation} \label{eq:3genp}
\bar{p} = \prod_{i=1}^{N} p_{i}^{q_{i}}
\end{equation}

According to the equations given above and equations (\ref{eq:2divdef}), (\ref{eq:2IDS}), if $\bar{p} = \frac{1}{\Theta}$, $D = S$, and $I = 0$. Similarly, if $\bar{p} > \frac{1}{\Theta}$, $D > S$, and $I > 0$. However, if $\bar{p} < \frac{1}{\Theta}$, $D < S$, and $I < 0$. Thus it appears that the information to direct a generic work can also both be positive or negative. This indicates that the purpose of information is to affect working trends, allowing us to recognise information as the capacity to direct works with preference.

\section{The Source of Information in the Physical World}
\begin{figure}[h]
	\centering
	\captionsetup{width=0.75\textwidth}
	\includegraphics[width=0.50\textwidth]{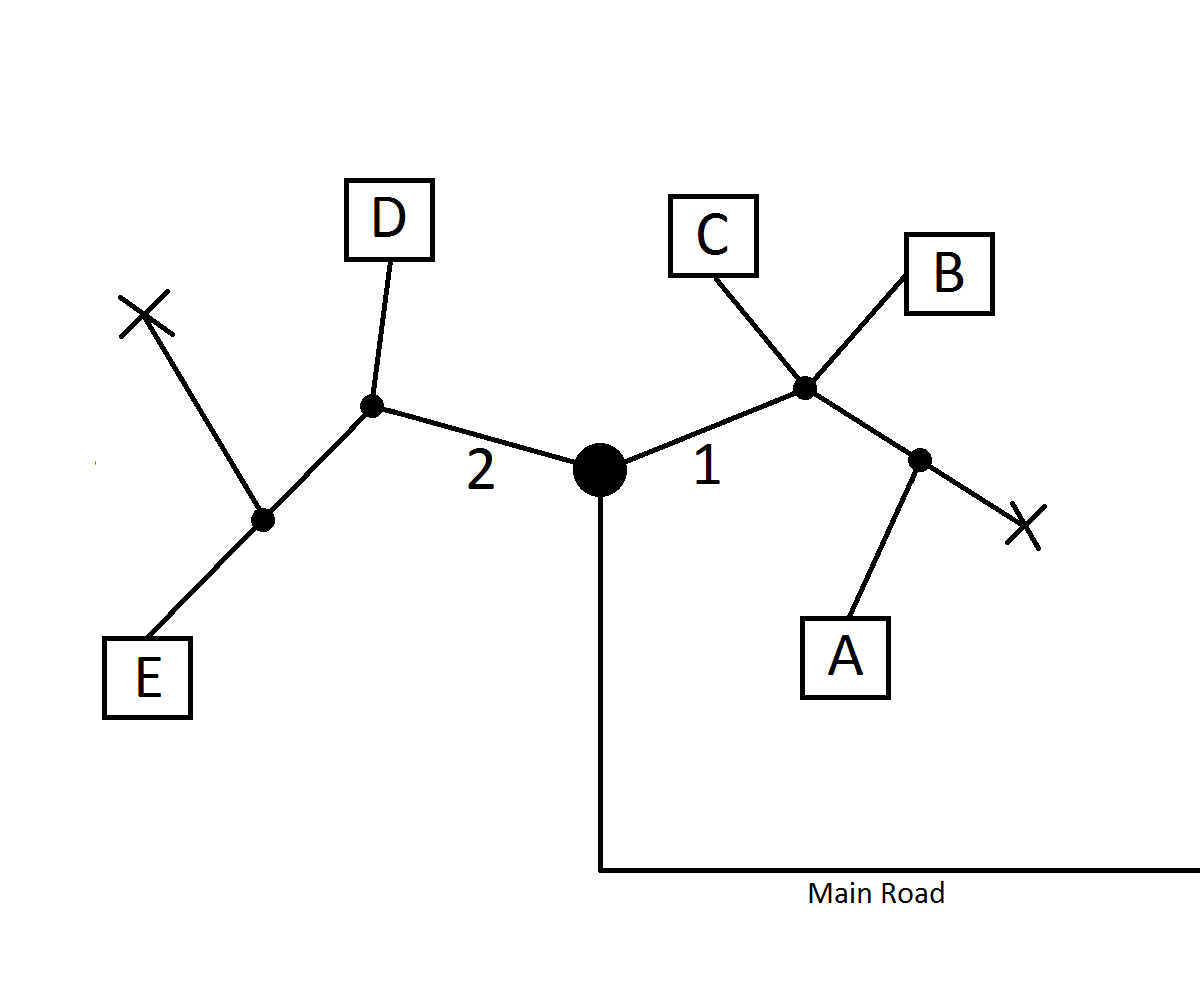}
	\caption{A diagram of Village Y, after the courier enters the village from the main road, there are 2 possible roads, that lead onto further junctions. After reaching each junction, the courier must decide which road to choose next. There are 5 residencies, A, B, C, D \& E and 2 branching roads lead to no residencies.}
	\label{fig:f2}
\end{figure}

In the previous sections, the information we discussed was always given by the worker. However, in our modern day understanding of information, we recognise that information can be provided by many different sources. To understand how other sources of information affect working processes, let us analyse Alan, a courier, delivering parcels in village Y. A diagram of village Y is depicted in figure \ref{fig:f2}.

In this example we can see that there are 7 working categories, and 5 working goals, same as in village X. Upon entering village Y, if Alan were to conduct his work by random trial without any information, we may deduce from figure \ref{fig:f2} that his probability of success of delivering a parcel to D would be $\frac{1}{4}$, and to A would be $\frac{1}{12}$. In comparison to delivering parcels to any residency in village X, where his probability of success was always $\frac{1}{7}$ when not directed by any information, it would appear that Alan's work in village Y is affected by some information. Upon closer inspection, it seems that the information comes from the uneven road layout structure in village Y. This example shows us that information may not only come from the worker himself, but also from other sources in the physical world. Maybe we can recognise anything that can provide information to create an uneven working trend as an information source.

Although Alan's work is affected by the road layout, he could also provide his own information to direct his work. Suppose Alan has some information which leads him to believe that road 2 has an 80\% chance of leading to D. Directed by this information, and the information given by the road layout, Alan’s probability of success delivering to D would be $\frac{2}{5}$. This indicates that the information Alan has can be used in conjunction with the information always given by the road layout, affecting his probability of success, suggesting information from different sources can simultaneously affect working processes.

In the above scenario however, the information Alan can provide, has no chance to affect his work by itself. Thus it seems that the amount of information given by Alan can only be represented as a relative amount of information, the quantity of which is only related to the change in probability of success, before and after the addition of Alan's information.

In a more general sense, suppose initially a specific work is directed by $I_{i}$ amount of information, with a probability of success $p_{i}$. After the introduction of an additional source of information, the probability of success becomes $p_{f}$, affected by a corresponding $I_{f}$ total amount of information. The relative amount of information given by the additional source would be $I_{r} = I_{f} - I_{i}$. According to equation (\ref{eq:1infodef}), if the introduction of an additional source of information does not change the working diversity we arrive at

\begin{equation} \label{eq:4relinf}
I_{r} = k \log \frac{p_{f}}{p_{i}}
\end{equation}

\noindent where $I_{r}$, $p_{i}$ and $p_{f}$ are as given above, and $k$ is a constant of proportionality. However, if the introduction of an additional information source changes the working diversity of the system, the amount of relative information would instead be given by

\begin{equation}
I_{r} = k \log \frac{\Theta_{f} p_{f}}{\Theta_{i} p_{i}}
\end{equation}

\noindent where $\Theta_{f}$ is the final total number of working categories, $\Theta_{i}$ is the initial total number of working categories and the rest are as given above.

Seemingly, any specific or generic work can be affected by a multitude of information sources originating from any structure in the physical world, however, the amount of information provided by each source should be viewed as a relative amount of information. Whether the relative information provided is positive, negative or zero depends on its subsequent effect on the workers probability of success.

\section{The Implication of Shannon's Entropy}

In the scenarios discussed in the previous sections, the information provided always originated from within the working system. However, sometimes an information source may be removed from the working system, and cannot directly affect any working process. If we wish for the external information source to take effect, we must somehow introduce its information into the system.

For example, a student wishes to gain more information about world war II to complete his history homework. If he does not have access to enough material containing the relevant information directly at hand, he may have to visit a library to borrow some books, or try to download supplementary materials from the internet. If we closely analyse his actions, we find that they are equivalent to bringing additional information sources into his working system. We may also realise that there are 2 very different ways of transferring such information sources. One method is to physically move the information source into the working system via transportation, and the other is to duplicate the original information source via communication.

If we consider the download process given above as a communications process, it would appear that such communications processes do not directly transfer the original information, but instead allow the communications recipient to duplicate the original information source. From this perspective, maybe we could recognise such communications processes as a working process involving duplication. Referring to the schematic of a general communication system given by Shannon\cite{shannon48}, the communications process can be seen as a process whereby, firstly the transmitter gains information about how the original information source is constructed; then the transmitter sends it to the receiver via the communications channel; lastly the receiver, directed by this information, replicates the original information source at the destination. Through this schematic, we understand why the information sent through the communications channel is irrelevant to the information given by the information source. 

At the destination, whether the original information is able to be fully gained is dependent on how well the original information source can be replicated. This indicates that the amount of information required to be sent through the communications channel is decided by the amount of information required by receiver to successfully duplicate the information source. The entropy discussed by Shannon pertaining to communications processes seems to be related to the working entropy of duplicating the original information source, which is precisely the amount of information required to be successfully sent through the communications channel. To explore the relationship between the entropy discussed by Shannon and the working entropy of duplication, let us look at the following example.

One day, Alice's friend sends her an e-mail about when her next tennis class was. However, by accident, she deletes the e-mail before she could read it. To precisely gain the information given by the e-mail, she would have to reproduce the original e-mail. Suppose the e-mail was written in a language consisting of $\Theta$ different symbols and is $N$ characters long. For the duplication work, both the number of working categories and working goals for reproducing any one symbol in the e-mail would be $\Theta$.

With no information to direct the reproduction work, the probability of success of reproducing each concurrent character in the e-mail would be $\frac{1}{\Theta}$. Therefore, the total working entropy, for the duplication of the entire e-mail would be $N \ln \Theta$. 

However, suppose Alice, via her understanding of the language, thinks the probability of occurrence of each symbol in the language is $(p_{1} , p_{2} , \ldots , p_{\Theta})$, with $\sum_{i}^{\Theta} p_{i} = 1$. By using this information to direct her replication work, she uses the above probabilities to dictate the probability at which each symbol is guessed for each subsequent character. Her resulting probability of success for guessing right each symbol is subsequently her probability of accurately replicating every symbol in the language, and would be the same as her supposed probability of occurrences. But the actual probability of occurrence of each symbol in this particular e-mail may be different, and is instead $(q_{1} , q_{2} , \ldots ,q_{N})$, with $\sum_{i=1}^{\Theta} q_{i} = 1$. Thus we may deduce the average working entropy for the reproduction of each individual character in this scenario, according to equation (\ref{eq:2divdef}), to be

\begin{equation} \label{eq:5charen}
S_{char} = - k \sum_{i=1}^{\Theta} q_{i} \log p_{i}
\end{equation}

\noindent $S_{char}$ is the average working entropy for the reproduction of each individual character, $k$ is a constant of proportionality and the rest are as given above.

We may notice that $p_{i}$ is an independent variable, only constrained by $\sum_{i}^{\Theta} p_{i} = 1$. By changing the values of this variable, the value of $S_{char}$ may decrease or increase. We found using Lagrange Multipliers that $S_{char}$ has a minimum value, when for each symbol $p_{i} = q_{i}$. The minimum value of $S_{char}$ is given by

\begin{equation} \label{eq:5charshan}
S_{min} = - k \sum_{i=1}^{\Theta} q_{i} \log q_{i}
\end{equation}

\noindent where $S_{min}$ is the minimum amount of working entropy for the reproduction of each character under this scheme, and the rest are as given above. Concerning the entire e-mail, her minimum amount of working entropy would be $- k N \sum q_{i} \log q_{i}$. It would appear that when Alice accurately knows the probability of occurrence of each symbol in the e-mail, and uses this information to direct her reproduction work, does she attain a minimum amount of working entropy, requiring the least amount of information to be re-sent through the communications channel for an accurate e-mail replication.

Shannon's analysis of communications systems\cite{shannon48} was based on the average probability of occurrence of any particular symbol for a language in general, and not that of any individual e-mail or telegram. However, since Shannon's expression for entropy $H = - k \sum p \log p$ has the same form as equation (\ref{eq:5charshan}), according to the above analysis, we may suppose Shannon's entropy has the same meaning as equation (\ref{eq:5charshan}) in representing the minimum working entropy. Correspondingly, the entropy he calculated in his work could be recognised as the minimum working entropy related to some replication work of any single character in a general communications process, when the symbol distribution of the used language is utilised to direct the work. By extension, if we apply these ideas to a more general working system, we may suppose Shannon's entropy as representing the minimum working entropy related to any guessing or replication work of the objects within that system, when directed by the distribution information of the system's objects. Naturally, in a communications system, Shannon's entropy would be the minimum amount of information required to be successfully sent through the communications channel. Similarly under such a scheme, when Shannon's entropy is equal to zero, the information still required to be sent is also zero, and the receiver has the maximum amount of information to direct the replication work.

From the above analysis, we may further understand why Shannon's entropy is irrelevant to the information given by the content of any message to be sent through communications. In Alice's example, the information given by the content of the e-mail was intended to direct Alice to her tennis class punctually. However, the information required to be sent through the communications channel, an amount given by Shannon's entropy was instead intended to direct Alice in replicating the original e-mail. Similarly, we also understand why there are some limitations in using Shannon's entropy to represent information in many situations\cite{brier05} \cite{sloman11} \cite{smith00} \cite{savage11}, as Shannon's entropy is not information, but instead is only equal to the amount of information still required to be gained in some specific circumstances.

\section{Thermodynamic Information}
In classical thermodynamics, we know that when a system is in thermodynamic equilibrium, there is a certain distribution of states of the microscopic particles. From quantum mechanics, we have come to understand that the states the microscopic particles can exist in are discrete. Since the total energy of the equilibrium system is fixed, there must also be a maximum energy any single particle can take, resulting in a finite discrete spectrum of states available for the particle states. However, as we are able to determine the distribution of particles amongst the different particle states, if we wish to make an accurate prediction as to the state of any single particle, it would appear from the previous section, that the working entropy of such work would be modelled by Shannon's entropy. 

In a classical thermodynamic system we also know that the distribution is not a flat distribution, thus we may suppose that the distribution provides some information to direct any prediction work. If we are also able to determine the total working diversity involved with such prediction work, by using equation (\ref{eq:2IDS}) we would be able to calculate how much information the distribution can provide. To find a relationship between the working information, diversity and entropy and the known macrostates of a thermodynamic system, let us explore an idealised system as given below.

Suppose the idealised thermodynamic system consists of a monatomic ideal gas at thermodynamic equilibrium held within a container. We know that there is always a certain energy distribution in such a system. As the container is of finite size, according to quantum mechanics we know that for a particle in a box of finite length $L$, and total volume $V$, relating $V = L^{3}$, in any single dimension, the energies of each particle must be quantised according to

\begin{equation} \label{eq:6enlevel}
\epsilon_n = \frac{n^{2} h^{2}}{8mV^{\frac{2}{3}}}
\end{equation}

\noindent where $V$ is the volume of the box, $m$ is the mass of the particle, $h$ is Planck's constant, $n$ is the principle quantum number with $n = 1, 2, 3, 4...$, and $\epsilon_n$ is the energy of each particle with the corresponding quantum number.

Since the total amount of heat energy in the system is fixed, the energy of each energy level a single particle can occupy must be less than or equal to the total amount of energy in the system, meaning the maximum quantum number is finite. If we consider this system to be a three dimensional system, we may suppose the energy levels must be axis invariant, meaning we may give for a container of arbitrary volume, following a density of states argument, a maximum quantum number of

\begin{equation} \label{eq:6maxn}
n_{m} = \bigg(\frac{8m V^\frac{2}{3} E}{h^{2}} \bigg)^{\frac{1}{2}}
\end{equation}

\noindent where $n_m$ is the maximum quantum number, $E$ is the total heat energy of the system, and the rest is given as above.

As particles occupying the same energy level may move in different directions in a three dimensional system, we must also consider energy level degeneracy. Thus the total number of momentum states any single particle can occupy in a thermodynamic system would be given by the sum of degeneracies of all energy levels. If we consider each single possible momentum state as corresponding to a particular working category in the case of doing some prediction work, the total number of working categories would be given by

\begin{equation} \label{eq:6sumdegen}
\Theta = \sum_{n=1}^{n_{m}} g_{n}
\end{equation}

\noindent where $\Theta$ is the number of working categories, $n_m$ is the maximum quantum number and $g_n$ is the degeneracy of each energy level with quantum number $n$. If $n_m$ is large, and by the density of states approximate $g_{n} = \pi n^{2}$, we can approximate $\Theta$ to give

\begin{equation} \label{eq:6thetasum}
\Theta = \frac{\pi}{3} n_{m}^{3}
\end{equation}

If we substitute (\ref{eq:6maxn}) into (\ref{eq:6thetasum}), we have

\begin{equation} \label{eq:6thetaE}
\Theta = \frac{\pi}{3} V \bigg(\frac{8 m E}{h^2}\bigg)^{\frac{3}{2}}
\end{equation}

We understand that at thermodynamic equilibrium, the total heat energy of the system may be given by $E = \frac{3}{2} N k_{B} T$, this gives us

\begin{equation} \label{eq:6theta}
\Theta = \frac{\pi}{3} V \bigg(\frac{12 k_B m N T}{h^{2}}\bigg)^{\frac{3}{2}}
\end{equation}

\noindent where $N$ is the total number of particles, $k_{B}$ is Boltzmann's constant and $T$ is the temperature of the system.

By using equation (\ref{eq:2divdef}), we may find the working diversity corresponding to predicting the momentum state of each single particle. However, if we wish to find the total working diversity for a system of $N$ particles, we would need to take $N \ln \Theta$. We use equation (\ref{eq:6theta}) to derive an expression for the working diversity of predicting all $N$ particle's momentum states, after re-arranging and ignoring a small remaining constant, we have

\begin{equation} \label{eq:6Ndiv}
D = N \ln V + \frac{3}{2} N \ln T + \frac{3}{2} N \ln N + \frac{3}{2} N \ln \frac{12 k_{B} m}{h^{2}}
\end{equation}

\noindent where $D$ is the working diversity, and the rest are as given above. From (\ref{eq:6theta}) and (\ref{eq:6Ndiv}), we see that $\Theta$ and $D$ are only related to the finite dynamical variables $V$, $T$, $N$ and $m$ for a classical non relativistic thermodynamic system, meaning the total working diversity and number of working categories of the system must also be finite.

As alluded to above, due to such a thermodynamic system's $\Theta$ being finite, we may use Shannon's entropy to calculate the minimum working entropy related to the prediction of any single particles momentum states. This would be given by

\begin{equation} \label{eq:6enthetasum}
S_{min} = \sum_{i=1}^{\Theta} - p_{i} \ln p_{i}
\end{equation}

\noindent where $S_{min}$ is the minimum working entropy, $\Theta$ the total number of working categories, and $p_{i}$ the probability of occurrence of a single particle in a given momentum state with $\sum p_{i} = 1$. Similarly, for the analysis of thermodynamic systems, we will henceforth be using natural logarithms corresponding to units of $nats$, dropping the constant $k$ as it will always be unity.

If we suppose that the probability of occurrence of each degenerate momentum state at any particular energy level is equal, and the probability of a particle being at a certain energy level is given by $q_{n}$ with a corresponding degeneracy of $g_{n}$, we have

\begin{equation} \label{eq:6ennsum}
S_{min} = \sum_{n=1}^{n_{m}} - q_{n} \ln \bigg( \frac{q_{n}}{g_{n}} \bigg)
\end{equation}

\noindent where $n_{m}$ is the maximum quantum number and the rest is as given above.

Similarly, for the system at thermodynamic equilibrium, we have $\sum \epsilon_{n} q_{n} = \frac{3}{2} k_{B} T$ and $\sum q_{n} = 1$. Therefore, using these equations, at thermodynamic equilibrium we may derive [Appendix]

\begin{equation} \label{eq:6enq}
q_{n} = \frac{g_{n} e^{- \frac{\epsilon_{n}}{k_{B}T}}}{\sum\limits_{i=n}^{n_{m}} g_{n} e^{- \frac{\epsilon_{n}}{k_{B}T}}}
\end{equation}

Substituting our expression for $q_{n}$ into equation (\ref{eq:6ennsum}), we may approximate an expression for $S_{min}$ [Appendix]

\begin{equation*}
S_{min} = \ln \bigg\lbrack \frac{V}{4} \bigg(\frac{8 \pi e m k_{B} T}{h^{2}} \bigg)^{\frac{3}{2}} \bigg\rbrack
\end{equation*}

Furthermore, if the system consists of $N$ particles, the total working entropy would be given by $S = N S_{min}$. By re-arranging the above expression for $S_{min}$ and ignoring a small remaining constant, we have

\begin{equation} \label{eq:6entropy}
S = N \ln V + \frac{3}{2} N \ln T + \frac{3}{2} N \ln \frac{12 k_{B} m}{h^{2}}
\end{equation}

\noindent where $S$ is the total minimum working entropy, and $b$ is an arbitrary constant, the rest are as given above. From this equation we may see that $S$ must have a finite value for the same reasons as the working diversity.

According to the relationship between information, diversity and entropy given by equation (\ref{eq:2IDS}), we may find the total amount of information that can be used to direct any prediction work, whilst only knowing the distribution of momentum states in the thermodynamic system, to be

\begin{equation} \label{eq:6info}
I = \frac{3}{2} N \ln N
\end{equation}

\noindent where $I$ is the information to direct our prediction work, and $N$ the number of particles. Interestingly, equation (\ref{eq:6info}) shows that in such a system, the amount of information is unrelated to the volume or temperature, and only related to the number of particles.

Obviously, the equations (\ref{eq:6Ndiv}), (\ref{eq:6entropy}) and (\ref{eq:6info}) show that our expressions for the working diversity, entropy and information are only related to the physical states of the system, and fairly irrelevant of the worker who does the predictions. This seems to suggest that our expressions for the working diversity, entropy and information can also be seen as extra macrostates to describe the thermodynamic system itself. Diversity would relate to the domain of momentum states in the thermodynamic system; entropy, the chaotic movement of the particles; information, the unevenness in the particle's momentum state distribution.

Furthermore, according to Boltzmann's expression for thermodynamic entropy, the thermodynamic entropy of the system as discussed above would be 

\begin{equation*}
S_{B} = k_{B} \ln W \, , \; \mbox{and} \; W = \frac{N! \prod_{n=1}^{n_{m}} g_{n}^{N_{n}}}{ \prod_{n=1}^{n_{m}} N_{n}!}
\end{equation*}

\noindent where $N$ is the total number of particles, $N_{n}$ is the number of particles in each energy level. By applying Stirling's approximation, and drawing the comparison that $q_{n} = \frac{N_{n}}{N}$, we find

\begin{equation*}
S_{B} = N k_{B} \sum_{n=1}^{n_{m}} - q_{n} \ln \bigg( \frac{q_{n}}{g_{n}} \bigg)
\end{equation*}

\noindent which when compared with equation (\ref{eq:6ennsum}), we easily derive

\begin{equation}
S_{B} = k_{B} S
\end{equation}

\noindent where $k_{B}$ is Boltzmann's constant. From the above equation, we see that to convert our expression for entropy to Boltzmann's entropy, we only need Boltzmann's constant, which acts as a factor of conversion, changing units from $nats$ to thermodynamic units. Therefore we may interpret Boltzmann's entropy as also representing the total uncertainty to predict which states all of the particles occupy. Similarly, this means that the entropy discussed in this section should also follow the second law of thermodynamics, and we may refer to the diversity, entropy and information we have discussed as thermodynamic diversity, entropy and information.

\section{Gibbs Paradox}
From the previous section, when a system consisting of a monatomic ideal gas is at thermodynamic equilibrium, with volume $V$, temperature $T$ and number of particles $N$, we are able to calculate the system's thermodynamic entropy. Suppose we split this system into 2 separate subsystems each with volume $\frac{1}{2}V$, if both subsystems are still at thermodynamic equilibrium with temperature $T$, then each subsystem must have $\frac{1}{2} N$ particles. According to equation (\ref{eq:6entropy}), we can find the change in total entropy, before and after splitting the system to be 

\begin{equation*}
\Delta S = 2 S' - S = - N \ln 2
\end{equation*}

\noindent where $\Delta S$ is the change in entropy, $S'$ is the entropy of each subsystem after splitting the system and $S$ is the entropy of the system before it was split. This result seems to show that the total entropy has decreased after splitting the system into two subsystems. If we still consider both subsystems as being part of the same isolated system, it would appear that the second law of thermodynamic has been violated, this is known as Gibbs paradox.

Previously, we used the working entropy of predicting a particle's state to understand a thermodynamic system's entropy. Maybe we could also take the same approach to help us understand Gibbs paradox. When the system separates into 2 subsystems, 2 different spatial states seem to emerge, as any particle can be in one subsystem or the other. From the perspective of doing prediction work, we must then not only predict what momentum state the particles are in, but also which spatial state. As the probability of any particle being in any single subsystem is $\frac{1}{2}$, the working entropy arising from predicting which spatial state each particle is in would be

\begin{equation} \label{eq:7Ss}
S_{s} = - N \bigg( \frac{1}{2} \ln \frac{1}{2} + \frac{1}{2} \ln \frac{1}{2}\bigg) = N \ln 2
\end{equation}

This suggests that the working entropy of such prediction work not only includes the thermodynamic entropy of each subsystem, but also the working entropy brought about by the spatial states of the system. It would appear the total working entropy after splitting the system should be $2S' + S_{s}$. This would yield $\Delta S = 0$, meaning the total working entropy would always remain the same no matter how we split the system.

Referring to equation (\ref{eq:7Ss}), it appears the spatial working entropy after the split is only related to the physical states of the system, meaning it may also be considered as a physical variable of the system itself. This implies the total working entropy discussed above of $2S' + S_{s}$ may be considered as the total entropy of the system. But we saw from the above example, before the split, all of the entropy arose from thermodynamic effects, which suggests that after splitting, some thermodynamic entropy transformed into what we may call spatial entropy. Since the spatial entropy $S_{s}$, is precisely the amount of entropy seemingly lost by splitting the system as presented in Gibbs paradox above, the entropy is not actually lost, and instead appears to have transformed from one type to another. As the total amount of entropy does not change, any physical processes which produce such transformations would be reversible, and do not violate the second law of thermodynamics.

From the previous discussion, we understood the diversity of a system is not dependent on information or entropy, but instead is only dependent on the state of the system itself. This suggests that for an isolated thermodynamic system, if its total volume, energy and number of particles are constant, no matter how its internal subsystems are arranged, the system's total diversity should remain unchanged. However, the distribution of particles amongst both spatial and momentum states seem to be governed by some corresponding information within the system. Naturally, the uncertainty in which particular state any particle is in would be represented by the corresponding type of entropy of the system. Following this idea, if a system remains isolated, according to equation (\ref{eq:2IDS}), different arrangements of the system's subsystems must have the following relationship

\begin{equation} \label{eq:7IDSsum}
D = \sum_{i=1}^{n} S_{i} + \sum_{i=1}^{n} I_{i} = \sum_{j=1}^{m} S_{j} + \sum_{j=1}^{m} I_{j}
\end{equation}

\noindent where $S_{i}$ \& $S_{j}$ are any single type of entropy, $I_{i}$ \& $I_{j}$ are the corresponding types of information, $n$ \& $m$ represent the different arrangements of the system, and $D$ is the total diversity of the entire system. By using this equation, we can calculate the amount of spatial information there is to govern the distribution of particles amongst the spatial states after the system was split in the example discussed above, which is $\frac{3}{2} N \ln 2$.

From the above discussion, we may see that when the diversity of the system is constant, any reversible process would yield

\begin{equation} \label{eq:7consinfoen}
\sum S_{i} = \sum S_{j} \; \; \mbox{and} \; \; \sum I_{i} = \sum I_{j}
\end{equation}

\noindent where $\sum S_{i}$ and $\sum I_{i}$ is the total entropy and information of the system in one arrangement, and $\sum S_{j}$ and $\sum I_{j}$, in another. Therefore, we find for any reversible process, there is conservation of total information and entropy.

\section{The Information to Conduct Thermodynamic Work}
We know that for a non-equilibrium thermodynamic system, as it tends towards equilibrium, it has the potential to do work. Since such a process involves entropy, from the previous sections analysis on the relationship between entropy and information, we may imagine this process must also involve information. To explore the changes in information and entropy in thermodynamic working processes, let us analyse the following example.

\begin{figure}[h]
	\centering
	\captionsetup{width=0.7\textwidth}
	\begin{subfigure}[b]{0.7\textwidth}
		\includegraphics[width=\textwidth]{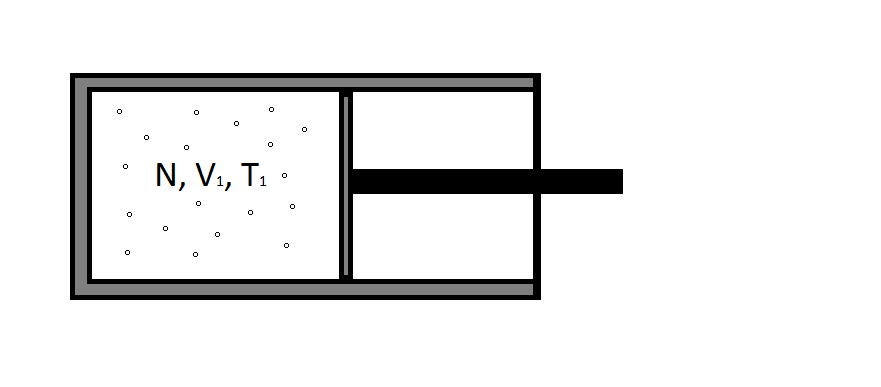}
		\captionsetup{width=0.85\textwidth}
		\caption{Piston in its initial state. All of the gas is in the left part of the piston in thermodynamic equilibrium. The gas consists of $N$ particles, volume $V_{1}$ and temperature $T_{1}$. The other part of the piston is a vacuum.}
		\label{fig:f3a}
	\end{subfigure}
	
	\begin{subfigure}[b]{0.7\textwidth}
		\includegraphics[width=\textwidth]{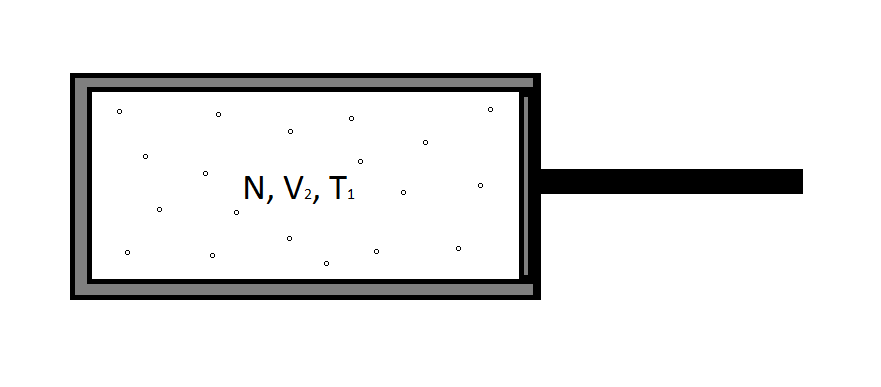}
		\captionsetup{width=0.85\textwidth}
		\caption{From the initial state, the piston undergoes free expansion to occupy all of the volume in the piston. When the gas reaches thermodynamic equilibrium, it has volume $V_{2}$, temperature $T_{1}$ and number of particles $N$.}
		\label{fig:f3b}
	\end{subfigure}
	
	\begin{subfigure}[b]{0.7\textwidth}
		\includegraphics[width=\textwidth]{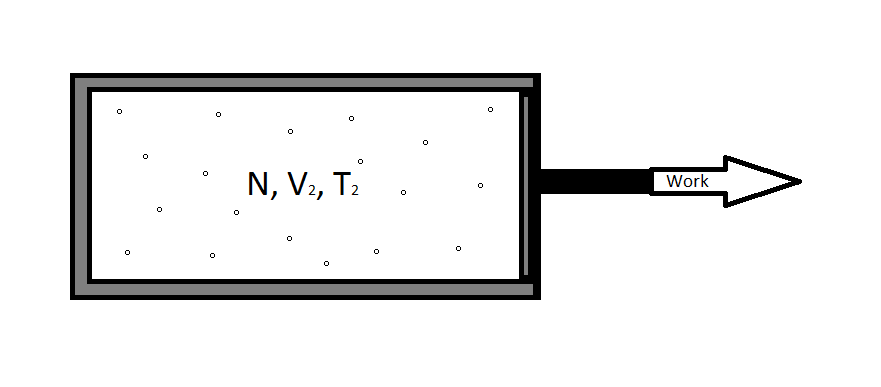}
		\captionsetup{width=0.85\textwidth}
		\caption{From the initial state, the piston undergoes reversible adiabatic expansion doing work, as depicted by the arrow. When the gas reaches thermodynamic equilibrium, it has volume $V_{2}$, temperature $T_{2}$ and number of particles $N$. }
		\label{fig:f3c}
	\end{subfigure}
	\caption{3 diagrams depicting different equilibrium states of a piston containing a monatomic ideal gas. The maximum volume the gas can occupy in the piston is $V_{2}$.}\label{fig:f3}
\end{figure}

Suppose there is an isolated piston, in its initial state, part of the piston with volume $V_{1}$ is filled with a monatomic ideal gas with $N$ particles and temperature $T_{1}$, whereas the other part is a vacuum, as depicted by figure \ref{fig:f3a}. Since the system is split into two subsystems, analogous to the two subsystems described in section 7, we must consider the systems total entropy as having been split up into 3 parts, the thermodynamic entropy of each individual subsystem, and the spatial entropy deriding from the subsystems arrangement within the system. Initially, as all of the energy and particles were in the part of the system with volume $V_{1}$, this subsystem must have thermodynamic entropy $S_{1}$ according to equation (\ref{eq:6entropy}) of

\begin{equation*}
S_{1} = N \ln V_{1} + \frac{3}{2} N \ln T_{1} + \frac{3}{2} N \ln \frac{12 k_{B} m}{h^{2}}
\end{equation*}

\noindent and thermodynamic information $I_{1}$ according to equation (\ref{eq:6info}) of

\begin{equation*}
I_{1} = \frac{3}{2} N \ln N
\end{equation*}

However, the other subsystem is a vacuum, containing no particles, its corresponding thermodynamic entropy $S_{0}$ and information $I_{0}$ would be zero. Since the probability of finding all of the particles within one subsystem is 1, and within the other subsystem is 0, from our preceding discussion, we understand that there would be zero initial spatial entropy $S_{s}$.

Since the piston is an isolated system and its diversity is only related to the states of the system, and not any particular arrangement, we may use equation (\ref{eq:6Ndiv}) to calculate the system's total diversity, which is

\begin{equation*}
D = N \ln V_{2} + \frac{3}{2} N \ln T_{1} + \frac{3}{2} N \ln N + \frac{3}{2} N \ln \frac{12 k_{B} m}{h^{2}}
\end{equation*}

This allows us to deduce from equation (\ref{eq:7IDSsum}), the amount of spatial information $I_{s}$ to be

\begin{equation} \label{eq:8insinfo}
I_{s} = D - (S_{1} + S_{0} + S_{s}) - (I_{1} + I_{0}) = N \ln \frac{V_{2}}{V_{1}}
\end{equation}

\noindent where $V_{1}$ is the volume of the system in its initial state, and $V_{2}$ is the maximum volume the gas can occupy in the system.
 
Suppose the system undergoes free adiabatic expansion, when the process has finished and has reached equilibrium, the system is no longer split into two subsystems (figure \ref{fig:f3b}), meaning our calculation of the total entropy and information no longer needs to be split into three parts. The residing total entropy of the system now only arise from thermodynamic affects, which is

\begin{equation*}
S_{2} = N \ln V_{2} + \frac{3}{2} N \ln T_{1} + \frac{3}{2} N \ln \frac{12 k_{B} m}{h^{2}}
\end{equation*}

\noindent and the residing total information of the system is also only thermodynamic information, which is

\begin{equation*}
I_{2} = \frac{3}{2} N \ln N
\end{equation*}

Comparing $S_{2}$ with $S_{1}$, we find the total amount of entropy in the system after undergoing free adiabatic expansion has increased, which we find to be

\begin{equation} \label{eq:8deltaS}
\Delta S = S_{2} - (S_{1} + S_{0} + S_{s}) = N \ln \frac{V_{2}}{V_{1}}
\end{equation}

Similarly, after expansion, there was no longer any spatial information left and the total amount of information changed by

\begin{equation} \label{eq:8deltaI}
\Delta I = I_{2} - (I_{1} + I_{0} + I_{s}) = - N \ln \frac{V_{2}}{V_{1}}
\end{equation}

Comparing the expressions given by equations (\ref{eq:8deltaS}) and (\ref{eq:8deltaI}), they quite clearly show us that the non-equilibrium spatial information initially present in the system completely transformed into thermodynamic entropy during the irreversible adiabatic free expansion, and no work is done, along with the system losing all potential to do any further work. As free adiabatic expansion is an irreversible process, we may see that for an irreversible process, there is

\begin{equation} \label{eq:8entoinfo}
\sum S_{i} - \sum S_{j} = - \bigg(\sum I_{i} - \sum I_{j} \bigg) > 0
\end{equation}

\noindent where $S_{i}$ and $I_{i}$ denote the systems entropy and information in its initial state; $S_{j}$ and $I_{j}$ denotes its final state. Comparing this equation to equation (\ref{eq:7consinfoen}), we understand that there is no longer any conservation of entropy or information for an irreversible process and instead, some information always transforms into entropy.

Conversely, we know that the system can also undergo reversible adiabatic expansion to do work. If the system were to undergo such a process, in its final state as depicted by figure \ref{fig:f3c}, we may still consider the system as being split into 2 subsystems. 1 subsystem contains both heat energy and all of the particles, and the other subsystem only consists of energy from work. We find the thermodynamic entropy in the subsystem with all of the particles to be

\begin{equation*}
S_{3} = N \ln V_{2} + \frac{3}{2} N \ln T_{2} + \frac{3}{2} N \ln \frac{12 k_{B} m}{h^{2}}
\end{equation*}

\noindent and thermodynamic information

\begin{equation*}
I_{3} = \frac{3}{2} N \ln N
\end{equation*}

Since there are no particles in the other subsystem, its thermodynamic entropy $S_{4}$ and information $I_{4}$ are zero. Similarly, the systems spatial entropy $S_{s}'$ is zero. Using equation (\ref{eq:7IDSsum}) we find that the spatial information present is

\begin{equation*}
I_{s}' = D - (S_{3} + S_{4} + S_{s2}) - (I_{3} + I_{4}) =\frac{3}{2} N \ln \frac{T_{1}}{T_{2}}
\end{equation*}

However, as $T V^{\frac{2}{3}} = \mbox{constant}$ for a reversible adiabatic process involving a monatomic gas, we find that $I_{s}' = I_{s}$ and $S_{3} = S_{1}$. This result shows us that in such a process there was no loss of spatial information, and no increase of thermodynamic entropy, despite the arrangement of the system's subsystems changing. By comparison with the scenario where the system underwent free adiabatic expansion, it appears that the reason the system still has the potential to do work after expansion is because the spatial information arising from the non-equilibrium states of the system still exist. The above result suggests to us that it is precisely this non-equilibrium spatial information which conducts thermodynamic work.

If this information can really conduct thermodynamic work, then it must seemingly represent some aspect of the system. According to equation (\ref{eq:6enlevel}), as a systems volume increases, the energy magnitude corresponding to each energy level decreases. In the above piston example, when it expanded from volume $V_{1}$ to $V_{2}$, each initial energy level must have lost some energy, which is

\begin{equation} \label{eq:8deltaelevel}
\Delta \epsilon_{n} = \frac{n^{2} h^{2}}{8m} \bigg( \frac{1}{V_{1}^{\frac{2}{3}}} - \frac{1}{V_{2}^{\frac{2}{3}}}\bigg)
\end{equation}

From this equation, it appears that if the system has the potential to expand from volume $V_{1}$ to $V_{2}$, the energy magnitude of every initial energy level must decrease, meaning every particle also has the potential to give out a certain amount of energy, a trait seemingly represented by the non-equilibrium spatial information of the system.

If the system undergoes adiabatic free expansion, from equation (\ref{eq:6maxn}), as the total energy of the gas does not change and $V$ increases, we can see that the total number of energy levels must increase. It would appear that in the process of adiabatic free expansion where entropy increases, the energy given out by the particles is redistributed into other particles which now occupy some new energy levels. Similarly, as for a reversible adiabatic expansion there is $T V^{\frac{2}{3}} = \mbox{constant}$, according to equation (\ref{eq:6maxn}), such a process would see no increase in number of energy levels. Thus it appears that all of the energy that is given out by the particles is instead used to do work. Seemingly if the energy given out by the particles is used to do work, the process is reversible and no non-equilibrium spatial information is lost. However, if the energy is instead redistributed into particles occupying new energy levels and the system's entropy increases, the process is irreversible.

As we know the probability of finding a particle at each initial energy level, and also how much energy each particle can give out, the maximum work the system can do must equal the total energy given out by all the particles in the system, so there is

\begin{equation} \label{eq:8workdeltaE}
W_{m} = N \sum_{n=1}^{n_{m}} q_{n} \Delta \epsilon_{n}
\end{equation}

\noindent where $W_{m}$ is the maximum reversible work done by the system, and the rest are as given previously. Substituting in equations (\ref{eq:6enq}) \& \ref{eq:8deltaelevel}, and using the same approximation method as used in the appendix to calculate the minimum entropy, we deduce that

\begin{equation} \label{eq:8workV}
W_{m} = - \frac{3}{2} N k_{B} T_{1} \bigg[ \bigg(\frac{V_{2}}{V_{1}}\bigg)^{-\frac{2}{3}} - 1 \bigg]
\end{equation}

\noindent where $T_{1}$ is the initial temperature of the gas, meaning the term $\frac{3}{2} N k_{B} T_{1}$ is equal to the total heat energy $E$ of the system. If we substitute equation (\ref{eq:8insinfo}) into (\ref{eq:8workV}), we have

\begin{equation} \label{eq:8workinfo}
W_{m} = - E \bigg[ \bigg(e^{\frac{I_{s}}{N}}\bigg)^{-\frac{2}{3}} - 1 \bigg] 
\end{equation}

Equation (\ref{eq:8workinfo}) clearly shows us that the system's non-equilibrium spatial information is intimately related to the corresponding maximum thermodynamic work that can be done. It appears that the more non-equilibrium spatial information there is, the greater the maximum thermodynamic work that can be done. If the system were to lose this information to entropy, the system also seems to lose its potential to do thermodynamic work despite still having thermodynamic energy.

\section{Discussion}

From our analysis of working processes, we have come to realise that the reason information is needed to conduct work is because any working system always has some working diversity. Correspondingly, it is due to this working diversity that we are often unable to attain a perfect energy efficiency when doing work, and the amount of information present determines the efficiency within the working system. The difference between the system's working diversity and information is subsequently represented by the working entropy. If a system has a non-zero value of working entropy, the corresponding work cannot be done with 100\% efficiency. We can see that the notion of any working process not only involving energy, but also diversity, information and entropy may easily be extended to all working systems.
	
In section 5, we discovered that when a system only has some statistical information, the working entropy of the system is given by Shannon's entropy. This is why Shannon's entropy is very successful in the analysis of thermodynamic systems at equilibrium as seen in section 6, because such systems can only provide statistical information relating to the statistical distribution of states. However, many systems are not limited to only providing some statistical information, which may be why Shannon's entropy faces difficulty in the analysis of such systems. From our perspective of information affecting working processes, our analysis does not appear to only be limited to communications systems or thermodynamic systems where there is some statistical information. This suggests that we may have the opportunity to find how information affects working processes in a much wider variety of scenarios embedded in differing disciplines.

For example, molecules involved in chemical reactions can interact in many different ways, but only when the relevant molecules interact in a certain way can a reaction occur. If we view a particular chemical reaction as a working goal, then there is only a certain probability of success of each interaction between molecules leading to a successful reaction. From the perspective of adding information to change the probabilities of success, maybe we can view catalysts in chemical reactions as information sources providing positive information, whereas inhibitors provide negative information.

Furthermore, in scientific research we often wish to investigate the cause and effects of certain phenomena. Since some research involves very complex conditions which lead to there being a very large working diversity, there may be lots of working entropy if we have very little knowledge concerning it. In such circumstances, we would hope to gain some information to direct our work by referring to previously published research. This allows us to understand why publications which provide negative results in their approaches can be seen as very informative, because they help us to eliminate certain non-conducive approaches in our future work, reducing working entropy. However, falsified or erroneous published results are harmful to future research, because they provide negative information.

We saw from the piston in section 8, an irreversible process results in information transforming into entropy. Since the second law of thermodynamics states that the entropy of an isolated system cannot decrease, we also realise that entropy cannot transform into information. This allows us to fathom that in an isolated system where diversity is constant, the system must inevitably reach a state of maximum entropy, attaining heat death. But from equation (\ref{eq:2deltaIDS}), even with the former constraints, it still appears possible for a system to reach a state further and further from equilibrium. Looking back at the piston, if the gas was permanently confined to an isolated system with volume $V_{1}$, ignoring the rest of the piston, upon reaching thermodynamic equilibrium, the system had no non-equilibrium information and its total diversity is only equal to the sum of its thermodynamic information and entropy. However, if the system were enlarged, and the gas was only held within a subsystem of the enlarged system, like in the piston in section 8, comparatively the system's diversity is now greater because the total volume of the entire system increased. Before the piston started to expand, the system now has a certain amount of non-equilibrium spatial information. Despite the non-equilibrium information having potential to transform into entropy, before such a transformation occurred, the system's total information was instead greater, and it would appear that the system was in a state further from equilibrium than before it was enlarged.

Maybe we can draw some comparisons between such an isolated piston and our universe. From this perspective, if we consider our universe as an isolated system with constant volume and total energy, it must inevitably reach heat death. However, if our universe is instead a perpetually expanding universe, we may suppose its diversity to be constantly increasing too. Like in the piston where an increase in diversity brought about some new non-equilibrium information, an increase in the universe's diversity could also bring about some new non-equilibrium information. If the rate at which this non-equilibrium information is created is greater than the rate at which information transforms into entropy in our universe, the total amount of information would increase and the universe would seemingly grow further and further away from equilibrium.

In section 8, maybe we can consider the non-equilibrium information that conducts the piston's thermodynamic work as originating from the spatial structure of the piston. After the piston did reversible thermodynamic work, we saw that there was no loss in information, yet the spatial structure of the piston had seemingly disappeared whilst a new structure had formed between the work energy outside of the piston and the piston itself. This seems to suggest that the information had been transferred from one structure to another. However, if we suppose that the work energy was next used to form a matter structure without increasing entropy, with respect to the entire system, it still does not appear that any non-equilibrium information has been lost. If the matter structure now has no further affect on the piston, we may imagine that the non-equilibrium information exists independently within the newly formed matter structure. As there is no loss in information to entropy, the above process should still be reversible. This seems to suggest that any initial non-equilibrium information can be transformed to be contained within different structures via certain working processes.

Following this train of thought, we can imagine that the non-equilibrium information which emerges from the increase in diversity of the universe may be able to transform into various different structures. Naturally, as this transformation takes place due to some working process, with an increase in the universe's diversity, the variety of structures could also increase. This could possibly result in an ever increasing variety of matter and structures in the universe, such as many different molecules and star formations, and may also have the chance to give rise to vastly more complex constructs.

We know that living organisms consist of a combination of complex molecules, which when acting together can seemingly resist collective decay. When part of a living organism decays, some of the organisms complex structures are damaged, we may consider such a process as one where the information contained within those structures are lost to entropy. To survive, the organism must remove the entropy from within itself and restore the information lost from the damaged structures by absorbing and transforming information from external sources. Naturally, this process must be faster than the rate at which the corresponding part of the organism decays. We may suppose that the reason living organisms consist of the complex molecules that they do, is due to the information given by those molecules acting together to enable any restoration processes being fast enough to resist total decay. If in our expanding universe, the total amount of information is really increasing alongside a growing variety of matter, given enough time, maybe in certain locations of the universe, the appropriate complex molecules have the opportunity to meet and conjoin, giving life a chance to spontaneously begin. Naturally, as life evolves from simple combinations to more and more complex organisms, it seems that the process of evolution is one whereby organisms gain more and more information. Following this train of thought, maybe we can argue that in our universe, life can spontaneously emerge and evolve without violating the second law of thermodynamics.

From our above analysis, information not only seems to exists in our brains, artificial constructs or organic materials, but instead seems to exist in every corner of our world. Similarly, no matter in which natural phenomena it is involved in, or in which form it exists, its role is always to conduct and affect working processes. Perhaps information should be seen as a fundamental scientific concept for us to use in understanding our world.

\newpage

\section*{Appendix} \label{App:Appendix}

From the example in section 6, when the system reaches thermodynamic equilibrium, we have

\begin{align}
& \sum_{n=1}^{n_{m}} - q_{n} \ln \bigg( \frac{q_{n}}{g_{n}} \bigg) = S_{min} \tag{A.1} \label{eq:A1} \\
& \sum_{n=1}^{n_{m}} q_{n} \epsilon_{n} = \frac{3}{2} k_{B} T \tag{A.2} \label{eq:A2} \\
& \sum_{n=1}^{n_{m}} q_{n} = 1 \tag{A.3} \label{eq:A3}
\end{align}

Similarly, at thermodynamic equilibrium, $S_{min}$ and $T$ are constant. If we find the derivatives of the above 3 equations, we have

\begin{align}
& \sum_{n=1}^{n_{m}} \bigg( \ln \frac{q_{n}}{g_{n}} \mathrm{d} q_{n} + \mathrm{d} q_{n} \bigg) = 0 \tag{A.4} \label{eq:A4} \\
& \sum_{n=1}^{n_{m}} \epsilon_{n} \mathrm{d} q_{n} = 0 \tag{A.5} \label{eq:A5} \\
& \sum_{n=1}^{n_{m}} \mathrm{d} q_{n} = 0 \tag{A.6} \label{eq:A6}
\end{align}

Substituting equation (\ref{eq:A6}) into (\ref{eq:A4}) we have

\begin{equation} \tag{A.7} \label{eq:A7}
\sum_{n=1}^{n_{m}} \ln \frac{q_{n}}{g_{n}} \mathrm{d} q_{n} = 0
\end{equation}

We may multiply equation (\ref{eq:A6}) by $\ln \alpha$ and equation (\ref{eq:A5}) by $\beta$, two lagrange multipliers. Summing these two new equations with equation (\ref{eq:A7}) we have

\begin{equation} \tag{A.8} \label{eq:A8}
\sum_{n=1}^{n_{m}} \bigg( \ln \frac{q_{n}}{g_{n}} + \beta \epsilon_{n} + \ln \alpha \bigg) dq_{n} = 0
\end{equation}

\noindent which must hold true for all $n$, giving

\begin{equation} \tag{A.9} \label{eq:A9}
\ln \frac{q_{n}}{g_{n}} + \beta \epsilon_{n} + \ln \alpha = 0
\end{equation}

\noindent re-arranging for $q_{n}$, we have

\begin{equation} \tag{A.10} \label{eq:A10}
q_{n} = \frac{g_{n}}{\alpha}  e^{- \beta \epsilon_{n}}
\end{equation}

\noindent substituting equation (\ref{eq:A10}) into equation (\ref{eq:A3}) we find

\begin{equation} \tag{A.11} \label{eq:A11}
\alpha = \sum_{n=1}^{n_{m}} g_{n} e^{- \beta \epsilon_{n}}
\end{equation}

\noindent combining equations (\ref{eq:A11}) and (\ref{eq:A10}), and substituting into equation (\ref{eq:A2}), we have

\begin{equation} \tag{A.12} \label{eq:A12}
\sum\limits_{n=1}^{n_{m}} \frac{\epsilon_{n} g_{n} e^{- \beta \epsilon_{n}}}{\sum\limits_{n=1}^{n_{m}} g_{n} e^{- \beta \epsilon_{n}}} = \frac{3}{2} k_{B} T
\end{equation}

From equation (\ref{eq:6enlevel}), if we let $\frac{h^{2}}{8 m V^{\frac{2}{3}}} = \gamma $, we have $\epsilon_{n} = \gamma n^{2}$. Similarly, from our discussion in section 6, we let $g_{n} = \pi n^{2}$. Introducing these expressions into equation (\ref{eq:A12}), we get

\begin{equation} \tag{A.13} \label{eq:A13}
\frac{\sum\limits_{n=1}^{n_{m}} \gamma n^{4} e^{- \beta \gamma n^{2}}}{\sum\limits_{n=1}^{n_{m}} n^{2} e^{- \beta \gamma n^{2}}} = \frac{3}{2} k_{B} T
\end{equation}

\noindent supposing our value of $n_{m}$ is very large, we may approximate our sum as an integral with limits from $0$ to $\infty$ and make the substitute of $n = x$, giving

\begin{equation} \tag{A.14} \label{eq:A14}
\frac{\int_{0}^{\infty} \gamma x^{4} e^{- \beta \gamma x^{2}} \mathrm{d} x}{\int_{0}^{\infty} x^{2} e^{- \beta \gamma x^{2}} \mathrm{d} x} = \frac{3}{2} k_{B} T
\end{equation}

\noindent using standard Gaussian integrals, we have

\begin{equation} \tag{A.15} \label{eq:A15}
\frac{\gamma \frac{3}{8} \sqrt{\frac{\pi}{(\beta \gamma)^{5} } } }{\frac{1}{4} \sqrt{ \frac{\pi}{(\beta \gamma)^{3} } } } = \frac{2}{3} k_{B} T
\end{equation} 

\noindent Simplifying equation (\ref{eq:A15}), we find that

\begin{equation} \tag{A.16} \label{eq:A16}
\beta = \frac{1}{k_{B} T}
\end{equation}

\noindent Combining equations (\ref{eq:A10}) and (\ref{eq:A11}), and substituting into it (\ref{eq:A16}), we gain an expression for $q_{n}$

\begin{equation} \tag{A.17} \label{eq:A17}
q_{n} = \frac{g_{n} e^{- \frac{\epsilon_{n}}{k_{B}T}}}{\sum\limits_{n=1}^{n_{m}} g_{n} e^{- \frac{\epsilon_{n}}{k_{B}T}}}
\end{equation}

\noindent Substituting equation (\ref{eq:A17}) into (\ref{eq:A1}), we obtain

\begin{equation} \tag{A.18} \label{eq:A18}
S_{min} = - \sum\limits_{n=1}^{n_{m}} \frac{g_{n} e^{- \frac{\epsilon_{n}}{k_{B}T}}}{\sum\limits_{n=1}^{n_{m}} g_{n} e^{- \frac{\epsilon_{n}}{k_{B}T}}}
\ln\Bigg({\frac{ e^{- \frac{\epsilon_{n}}{k_{B}T}}}{\sum\limits_{n=1}^{n_{m}} g_{n} e^{- \frac{\epsilon_{n}}{k_{B}T}}}} \Bigg)
\end{equation}

\noindent which upon re-arranging yields

\begin{equation} \tag{A.19} \label{eq:A19}
S_{min} = \frac{ \sum\limits_{n=1}^{n_{m}} \frac{\epsilon_{n}}{k_{B}T} g_{n} e^{- \frac{\epsilon_{n}}{k_{B}T}}}{\sum\limits_{n=1}^{n_{m}} g_{n} e^{- \frac{\epsilon_{n}}{k_{B}T}}} + \ln \bigg( \sum\limits_{n=1}^{n_{m}} g_{n} e^{- \frac{\epsilon_{n}}{k_{B}T}}\bigg)
\end{equation}

\noindent again, if we let $g_{n} = \pi n^{2}$ and $\epsilon_{n} = \gamma n^{2}$ as given previously, approximate our sum to an integral for large $n_{m}$ and use the substitution of $n = x$, we have

\begin{equation} \tag{A.20} \label{eq:A20}
S_{min} = \frac{\int_{0}^{\infty} \frac{\gamma x^{2}}{k_{B}T} \pi x^{2} e^{- \frac{\gamma x^{2}}{k_{B}T}} \mathrm{d} x}{\int_{0}^{\infty} \pi x^{2} e^{- \frac{\gamma x^{2}}{k_{B}T}} \mathrm{d} x } + \ln \bigg( \int_{0}^{\infty} \pi x^{2} e^{- \frac{\gamma x^{2}}{k_{B}T}} \mathrm{d} x \bigg)
\end{equation}

\noindent which we can solve using standard Gaussian integrals, upon simplifying to give

\begin{equation} \tag{A.21} \label{eq:A21}
S_{min} = \ln \bigg\lbrack \frac{V}{4} \bigg(\frac{8 \pi e m k_{B} T}{h^{2}} \bigg)^{\frac{3}{2}} \bigg\rbrack
\end{equation}

\noindent where $V$ is the volume of the container, $T$ is the temperature, $m$ is the mass of each particle, $k_{B}$ is Boltzmann's constant, $h$ is Planck's constant, $e$ is the natural number, and $S_{min}$ is the minimum working entropy.

\bibliography{citations}

\newpage

\section*{Figures}

\end{document}